\documentclass[pra,twocolumn,showpacs]{revtex4}
\usepackage{amsmath,amssymb,mathrsfs}
\usepackage{psfrag}
\usepackage{graphicx}
\usepackage{graphics}
\usepackage{epsfig}
\usepackage{bm}
\usepackage{color}
\usepackage{verbatim,color,ulem}

\newcommand{\beq}{\begin{equation}}
\newcommand{\eeq}{\end{equation}}
\newcommand{\beqa}{\begin{eqnarray}}
\newcommand{\eeqa}{\end{eqnarray}}
\newcommand{\ba}{\begin{array}}
\newcommand{\ea}{\end{array}}

\begin{document}

\title{Discrete bright solitons in Bose-Einstein condensates \\ 
and dimensional reduction in quantum field theory}
\author{Luca Salasnich} 
\affiliation{
$^{1}$Department of Physics and Astronomy ``Galileo Galilei'' and CNISM, 
Universita di Padova, via Marzolo 8, 35131 Padova, Italy, \\
$^{2}$Istituto Nazionale di Ottica (INO) del Consiglio Nazionale delle
Ricerche (CNR), Sezione di Sesto Fiorentino, via Nello Carrara, 50019
Sesto Fiorentino, Italy \\
\email{luca.salasnich@unipd.it}}

\begin{abstract} 
We first review the derivation of an 
effective one-dimensional (1D) discrete nonpolynomial Schr\"odinger equation 
from the continuous 3D Gross-Pitaevskii equation 
with transverse harmonic confinement and axial periodic potential. 
Then we study the bright solitons obtained from 
this discrete nonpolynomial equation 
showing that they give rise to the collapse 
of the condensate above a critical attractive strength. 
We also investigate the dimensional reduction 
of a bosonic quantum field theory, deriving an effective 
1D nonpolynomial Heisenberg equation 
from the 3D Heisenberg equation of the continuous 
bosonic field operator under the action of transverse harmonic 
confinement. Moreover, by taking into account the presence of an axial 
periodic potential we find a generalized Bose-Hubbard model 
which reduces to the familiar 1D Bose-Hubbard Hamiltonian 
only if a strong inequality is satisfied. Remarkably, in the 
absence of axial periodic potential our 1D 
nonpolynomial Heisenberg equation gives the generalized 
Lieb-Liniger theory we obtained some years ago. 
\end{abstract}


\maketitle
 
\section{Introduction}

Ultracold bosonic gases in reduced dimensionality are an ideal platform 
for probing many-body phenomena ~\cite{giamarchi,cazalilla2011}. 
In particular, the use of optical lattices has allowed the experimental 
realization  \cite{bloch} of the well-known 
Bose-Hubbard Hamiltonian \cite{book-lattice} with dilute and 
ultracold alkali-metal atoms. 
This achievement has been of tremendous impact on several 
communities \cite{book-lattice}, and in particular on theoreticians 
and mathematicians working with discrete nonlinear Schr\"odinger 
equations \cite{kevre}. 

The three-dimensional (3D) Gross-Pitaevskii equation, a cubic 
nonlinear Schr\"odinger equations which accurately 
describes a Bose-Einstein condensate (BEC) made of dilute and 
ultracold atoms \cite{book-bose}, 
is usually analyzed in the case of repulsive interaction strength
which corresponds to a positive inter-atomic s-wave scattering length 
\cite{bloch_review}. Indeed, a negative s-wave 
scattering length implies an attractive interaction strength 
which may bring to the collapse \cite{book-bose}
due to the shrink of the transverse width of a realistic quasi-1D bosonic 
cloud \cite{sala-npse,sala-dnpse1,sala-dnpse2}. Nevertheless, 
in certain regimes of interaction 
both continuous and discrete 3D Gross-Pitaevskii equations 
predict the existence of 
meta-stable configurations which are usually called continuous 
and discrete bright solitons \cite{sala-npse,sala-dnpse1,sala-dnpse2}. 
We remark that continuous bright solitons have been observed 
in various experiments \cite{exp-solo1,exp-solo2,exp-solo3,exp-solo4} 
involving  attractive bosons of $^7$Li and $^{85}$Rb vapors. Instead, 
discrete (gap) bright solitons in quasi-1D optical lattices 
have been observed \cite{exp-gap} 
only with repulsive bosons made of $^{87}$Rb atoms. 

In the first part we discuss an 
effective one-dimensional discrete nonpolynomial Schr\"odinger equation 
obtained from the continuous 3D Gross-Pitaevskii equation 
with transverse harmonic confinement and axial periodic potential 
\cite{sala-dnpse1,sala-dnpse2}. We show that this 1D discrete 
nonpolynomial Schr\"odinger 
equation reduces to the 1D discrete Gross-Pitaevskii equation 
only in the weak-coupling regime and we compare the bright soliton of the 
discrete nonpolynomial Schr\"odinger equation 
bright solitons with the bright solitons of the 
discrete Gross-Pitaevskii equation. 

In the second part, we investigate the dimensional reduction 
of a bosonic quantum field theory, deriving an effective 
1D nonpolynomial Heisenberg equation 
from the 3D Heisenberg equation of the 
bosonic field operator under the action of transverse harmonic 
confinement. In particular, we prove that the discrete version of this 1D 
nonpolynomial Heisenberg equation becomes the 1D discrete 
nonpolynomial Schr\"odinger equation only assuming that the 
quantum many-body state of the system is a Glauber coherent state. 
As a by-product, we also obtain a reliable 
generalizaton of the Lieb-Liniger theory for a quasi-1D uniform 
Bose gas \cite{sala-lieb1}. 

It is important to stress that some years ago we used this generalized 
Lieb-Liniger theory (but in the absence of axial lattice) to analyze 
the transition from a 3D Bose-Einstein condensate 
to the 1D Tonks-Girardeau gas, showing that the sound velocity and 
the frequency of the lowest compressional mode give a clear signature 
of the regime involved \cite{sala-lieb1}. In Ref. \cite{sala-lieb1} we 
studied also the case of negative scattering length deriving 
the phase diagram of the Bose gas (uniform, single soliton, multi soliton 
and collapsed) in toroidal confinement. Quite remarkably, 
the experimental data on a Tonks-Girardeau gas of $^{87}$Rb atoms 
of Kinoshita, Wenger, and Weiss \cite{kinoshita} are compatible with the 
one-dimensional theory of Lieb, Seiringer and Yngvason \cite{lieb} 
but are better described by our theory that takes into account 
variations in the transverse width of the atomic cloud \cite{sala-lieb2}. 
In Ref. \cite{sala-lieb2}, by using our generalized theory we investigated 
also the free axial expansion of the $^{87}$Rb gas in different regimes: 
Tonks-Girardeau gas, one-dimensional Bose-Einstein condensate 
and three-dimensional Bose-Einstein condensate. 

\section{BEC in a quasi-1D optical lattice}

We consider a dilute BEC confined 
in the $z$ direction by a { generic axial potential} $V(z)$ 
and in the plane $(x,y)$ by the { transverse harmonic potential}
\beq
U(x,y) = {m\over 2} \omega_{\bot}^2
\left(x^2 + y^2 \right) \; . 
\eeq
The characteristic harmonic length is given by 
\beq 
a_{\bot}=\sqrt{\hbar\over m\omega_{\bot}} \; , 
\eeq
and, for simplicity, we choose $a_{\bot}$ and $\omega_{\bot}^{-1}$, 
as length and time units, and  $\hbar \omega_{\bot }$ as energy unit. 
In the remaining part of this chapter we use non-dimensional variabiles. 

We assume that the system made of fully condensed Bose atoms 
is well described by the 3D 
Gross-Pitaevskii equation, and in scaled units it reads
\beqa
i{\frac{\partial}{\partial t}}\psi(\mathbf{r},t) &=&
\Big[ -{\frac{1}{2}}\nabla^{2}
+ {1\over 2} \left( x^2 + y^2 \right)
+ V(z) 
\nonumber 
\\
&+& 
2\pi g |\psi(\mathbf{r},t) |^{2}\Big] \psi(\mathbf{r},t) \; , 
\label{3dgpe}
\eeqa
where $\psi(\mathbf{r},t)$ is the macroscopic wave function of the
BEC normalized to the total number $N$ of atoms 
and $g= 2a_s/a_{\bot}$ with $a_s$ the s-wave scattering length of the
inter-atomic potential. 
In addition, we suppose that { the axial potential is the 
combination of periodic and harmonic potentials}, i.e. 
\beq 
V(z) = V_0 \cos{(2 k z)} + {1\over 2} \lambda^2 z^2 \; .  
\label{periodic}
\eeq
This potential models the { quasi-1D optical lattice} 
produced in experiments with 
Bose-Einstein condensates by using counter-propagating 
laser beams \cite{morsch}. 
Here $\lambda \ll 1$ models a weak axial harmonic confinement. 

\subsection{Axial discretization of the 3D Gross-Pitaevskii equation } 

We now perform a discretization of the 3D Gross-Pitaevskii equation 
along the $z$ 
axis due to the presence on the periodic potential. 
In particular we set 
\beq
\psi(\mathbf{r},t) = \sum_n \phi_n(x,y,t) \ W_n(z) \; , 
\label{zanno}
\eeq
where $W_n(z)$ is the { Wannier function} maximally localized 
at the $n$-th minimum of the axial periodic potential. 
This tight-binding ansatz is reliable in the case of a deep optical 
lattice \cite{morsch}.

We insert this ansatz into Eq. (\ref{3dgpe}), multiply 
the resulting equation by $W_n^*(z)$ and integrate over $z$ variable. 
In this way we get 
\beqa
i{\partial\over \partial t} \phi_n &=& 
\Big[-\frac{1}{2}\nabla_{\bot}^2 + 
\frac{1}{2} \left( x^{2}+y^{2}\right) 
+ \epsilon_n \Big] \phi_n 
\nonumber 
\\
&-& J \left( \phi_{n+1}+\phi_{n-1}\right) + 
2\pi U \left\vert \phi_n \right\vert^{2} \phi_n 
\: , 
\label{array2D}
\eeqa
where the parameters $\epsilon$, $J$ and $U$ are given by 
\beq
\epsilon_n = \int W_n^*(z) \left[ -{1\over 2} {\partial^2\over \partial z^2} 
+ V(z) \right] W_n(z) \ dz  \; , 
\label{maroni1}
\eeq
\beq
J = - \int W_{n+1}^*(z) \left[ -{1\over 2}
{\partial^2\over \partial z^2} 
+  V(z) \right] W_n(z) \ dz  \; , 
\label{maroni2}
\eeq
\beq 
U = g \int |W_n(z)|^4 \ dz  \; . 
\label{maroni3}
\eeq
The parameters $J$ and $U$ are practically independent 
on the site index $n$ and in the tight-binding regime $J>0$. 

\subsection{Transverse dimensional reduction of the 3D discrete 
Gross-Pitaevskii equation}

To further simplify the problem we set \cite{sala-dnpse1,sala-dnpse2}
\beq
\phi_n(x,y) = 
{1\over \pi^{1/2} \sigma_n(t)}
\exp{\left[ - \left( {x^2+y^2\over 2\sigma_n(t)^2} 
\right) \right] }\,f_n(t) \; , 
\label{assume}
\eeq
where $\sigma_n(t)$ and $f_n(t)$, which account
for { discrete transverse width} and { discrete axial wave function}, 
are the effective generalized coordinates 
to be determined variationally. In Ref. \cite{sala-gaussian} 
there is a detailed discussion of the variational approach 
with time-dependent Gaussian trial wave-functions 
for the study of Bose-Einstein condensates. 

We insert this ansatz into the Lagrangian density associated 
to Eq. (\ref{array2D}) and integrate over $x$ and $y$ variables. 
In this way we obtain an effective Lagrangian for the fields 
$f_n(t)$ and $\sigma_n(t)$. 

The Euler-Lagrange equation of the effective Lagrangian 
with respect to $f_n^*$ is 
\beqa
i {\partial\over \partial t} f_n &=& 
\left[ {1\over 2} \left( {1\over \sigma_n^2} +  
\sigma_n^2 \right) + \epsilon_n \right] f_n 
- J \, \left( f_{n+1}+f_{n-1} \right) 
\nonumber 
\\
&+& {U \over \sigma_n^2} |f_n|^2 f_n \; . 
\label{e1}
\eeqa
while the Euler-Lagrange equation with respect to $\sigma_n$ gives 
\beq 
\sigma_n^4 = 1 + U |f_n|^2  \; . 
\label{sig1}
\eeq
Inserting Eq. (\ref{sig1}) into Eq. (\ref{e1}) we finally get 
\beq 
i {\partial\over \partial t} f_n = \epsilon_n \, f_n
- J \, \left( f_{n+1}+f_{n-1} \right) + 
\frac{1+{(3/2)}U|f_n|^{2}}{\sqrt{1+U|f_n|^{2}}} f_n \; ,  
\label{dnpse}
\eeq
that is the 1D { discrete nonpolynomial 
Schr\"odinger equation}, 
describing the BEC under a transverse anisotropic 
harmonic confinement and an axial optical lattice 
\cite{sala-dnpse1,sala-dnpse2}.

\begin{figure}[t] 
\begin{center}
\centerline{\epsfig{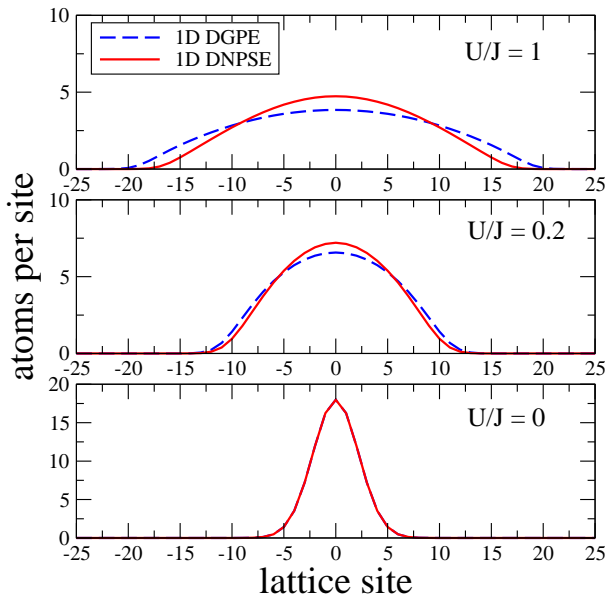}}
\end{center}
\vskip -0.5cm
\caption{Weakly repulsive bosons in the optical lattice. 
Axial density profile (atoms per site) of $N=100$ repulsive bosonic atoms  
in a optical lattice 
with a super-imposed harmonic potential ($\lambda=0.1$). 
The three panels correspond 
(from bottom to top) to increasing values of the adimensional on-site 
interaction strength $U/J: 0, 0.2, 1$. 
Solid lines: results obtained by using the 
1D discrete nonpolynomial Schrodinger equation (DNPSE); 
dashed lines: results obtained by using the 1D discrete Gross-Pitaevskii 
equation (DGPE). In the lower panel ($U/J=0$) the two curves 
are superimposed.}
\label{fig1}
\end{figure}

\begin{figure}[t] 
\begin{center}
\centerline{\epsfig{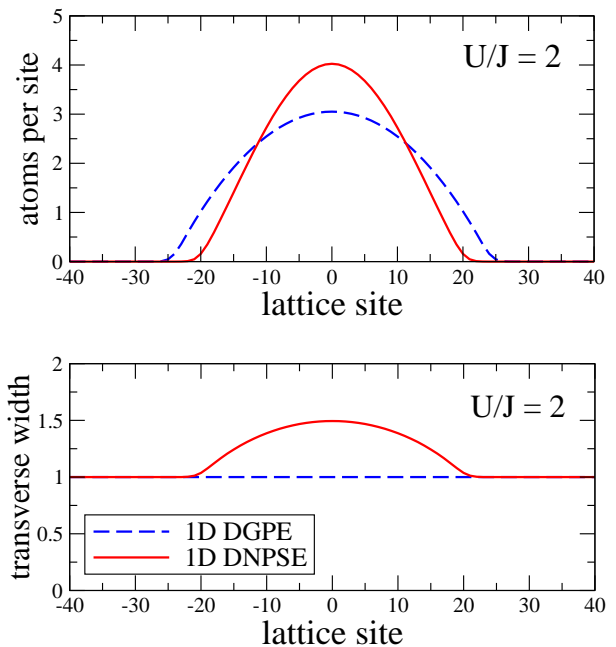}}
\end{center}
\vskip -0.5cm
\caption{Strongly repulsive bosons in the optical lattice ($U/J=2$). 
Upper panel: Axial density profile (atoms per site) of $N=100$ repulsive 
bosonic atoms in a optical lattice with a 
super-imposed harmonic potential ($\lambda=0.1$). 
Lower panel: the transverse width of the bosonic cloud. 
Solid lines: results obtained by using the 
1D discrete nonpolynomial Schrodinger equation (DNPSE); 
dashed lines: results obtained by using the 1D discrete Gross-Pitaevskii 
equation (DGPE).}
\label{fig2}
\end{figure}

The 1D discrete nonpolynomial Schr\"odinger equation reduces to the familiar 
1D discrete Gross-Pitaevskii equation 
\beq 
i {\partial\over \partial t} f_n = \epsilon_n \, f_n 
- J \, \left( f_{n+1} +f_{n-1} \right) + U|f_n|^{2} f_n \;  
\eeq
in the weak-coupling limit $|U| |f_n|^2\ll 1$, where $U$ can be 
both positive and negative. On the contrary, 
1D discrete nonpolynomial Schr\"odinger equation becomes a 1D quadratic 
discrete nonlinear Schr\"odinger equation 
\beq 
i {\partial\over \partial t} f_n = \epsilon_n \, f_n 
- J \, \left( f_{n+1} +f_{n-1} \right) + (3/2)\sqrt{U} |f_n| f_n \;  
\eeq
in the strong-coupling limit $U|f_n|^2\gg 1$, where $U>0$.  

\subsection{Numerical results}

We have solved numerically both 1D discrete nonpolynomial 
Schr\"odinger equation and 1D discrete Gross-Pitaevskii equation by using a 
{ Crank-Nicolson predictor-corrector algorithm} with imaginary time 
\cite{sala-numerics} to get the ground-state of the system. 

In Fig. \ref{fig1} and \ref{fig2} 
we report our results obtained with $N=100$ atoms 
in a quasi-1D optical lattice with weak axial harmonic confinement: 
$\lambda=0.1$. 
The plots are shown for different values of the { repulsive}  
on-site interaction strength $U$: $U>0$.  Note that in the experiments 
$U$ can be tuned by using the technique of Feshbach 
resonances \cite{book-lattice,book-bose,morsch}. 

In Fig. \ref{fig1} we plot the axial density profile $|f_n|^2$ 
of weakly repulsive bosons in a optical lattice 
with a super-imposed harmonic potential. As described in the caption, 
the three panels correspond 
(from bottom to top) to increasing values of the on-site 
interation strength $U$. Fig. \ref{fig1} clearly shows that the 
results (solid lines) obtained by using the 
1D discrete nonpolynomial Schrodinger equation  
strongly differ with respect to the ones (dashed lines) 
obtained by using the 1D discrete Gross-Pitaevskii 
equation by increasing the on-site interaction. 
This effect is better shown in the upper panel 
of Fig.  \ref{fig2}, where we plot the axial density profile 
for a large value ($U/J=2$) of the on-site interaction. 
In the lower panel of Fig.  \ref{fig2} we report 
the transverse width $\sigma_i$ of the bosonic cloud as a 
function of the lattice site $n$. As expected, $\sigma_i$ 
strongly deviates from $1$ (i.e. $a_{\bot}$ is dimensional units) where 
the axial density $|f_n|^2$ is large. 

\begin{figure}[t] 
\begin{center}
\centerline{\epsfig{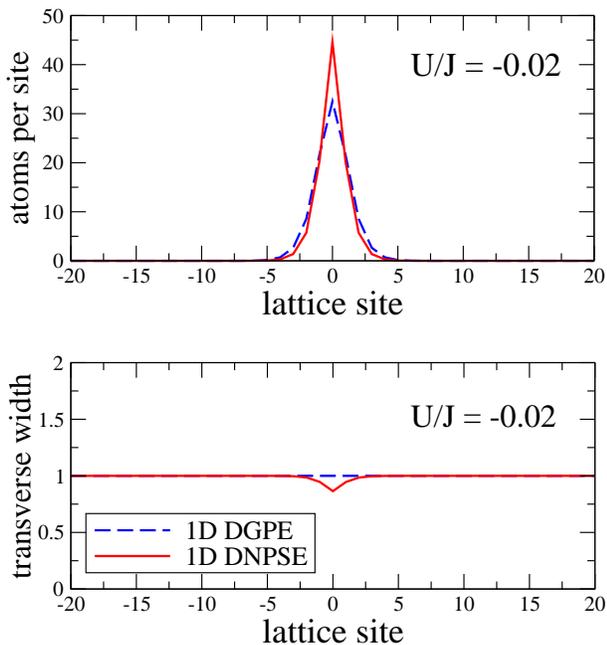}}
\end{center}
\vskip -0.5cm
\caption{Weakly attractive bosons in the optical lattice ($U/J=-0.02$). 
Upper panel: Axial density profile (atoms per site) 
of $N=100$ attractive bosonic atoms in a optical lattice with a 
super-imposed harmonic potential ($\lambda=0.1$). 
Lower panel: the transverse width of the bosonic cloud. 
Solid lines: results obtained by using the 
1D discrete nonpolynomial Schrodinger equation (DNPSE); 
dashed lines: results obtained by using the 1D discrete Gross-Pitaevskii 
equation (DGPE).}
\label{fig3}
\end{figure}

\begin{figure}[t]
\begin{center}
\centerline{\epsfig{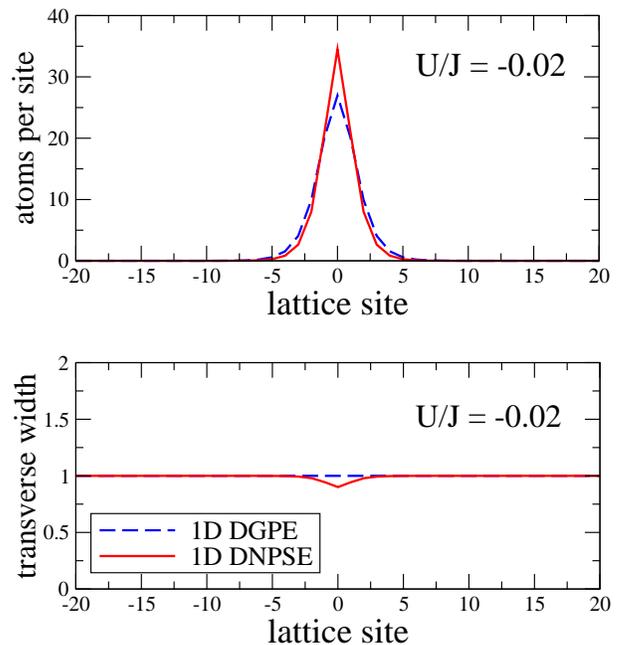}}
\end{center}
\vskip -0.5cm
\caption{Weakly attractive bosons in the optical 
lattice ($U/J=-0.02$) without the super-imposed harmonic 
confinement ($\lambda=0$). 
Upper panel: Axial density profile (atoms per site) of $N=100$ attractive 
bosonic atoms in a optical lattice. 
Lower panel: the transverse width of the bosonic cloud. 
Solid lines: results obtained by using the 
1D discrete nonpolynomial Schrodinger equation (DNPSE); 
dashed lines: results obtained by using the 1D discrete Gross-Pitaevskii 
equation (DGPE).}
\label{fig4}
\end{figure}

Now we show the results obtained again with $N=100$ atoms 
in a quasi-1D optical lattice but with an attractive  
on-site interaction strength $U$: $U<0$. 
In the attractive case the ground-state is self-localized 
and it exists also in the absence ($\lambda=0$) of the axial 
harmonic potential: it is the discrete bright soliton. 
In Fig. \ref{fig3} we plot the axial density profile $|f_n|^2$ 
in the presence of the super-imposed axial harmonic potential 
($\lambda = 0.1$) and in Fig. \ref{fig4} in the absence 
of the super-imposed axial harmonic potential ($\lambda = 0$) 
choosing $U/J=-0.02$. The two figures show that that the 
density profiles with and without axial harmonic 
potential are practically the same. In the figures 
there is also the comparison between 1D nonpolynomial Schr\"odinger 
equation (solid lines) and 1D Gross-PItaevskii equation (dashed lines). 

\subsection{Collapse of the discrete bright soliton} 

\begin{figure}[t] 
\begin{center}
\centerline{\epsfig{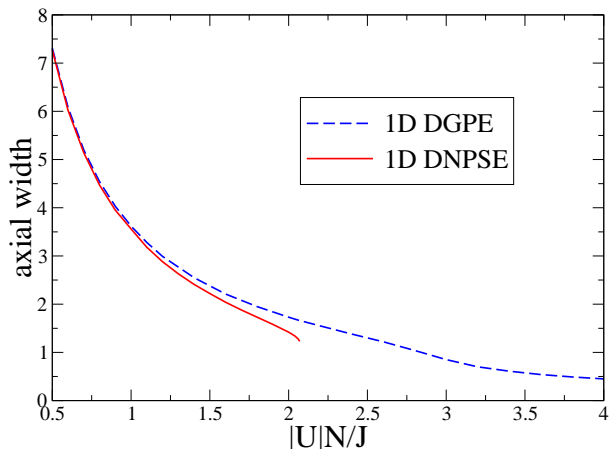}}
\end{center}
\vskip -0.5cm
\caption{Weakly attractive bosons in the optical 
lattice without the super-imposed harmonic confinement 
($\lambda=0$). Transverse width of the bosonic cloud of $N=100$ atoms 
as a function of the 
effective interaction strength $|U|N/J$. 
Solid lines: results obtained by using the 
1D discrete nonpolynomial Schrodinger equation (DNPSE); 
dashed lines: results obtained by using the 1D discrete Gross-Pitaevskii 
equation (DGPE).}
\label{fig5}
\end{figure}

In Fig. \ref{fig5} we report the axial width of the bright soliton 
as a fuction of the (attractive) on-site interaction. 
As expected, for a small on-site interaction strength the axial width 
is extremely large and 1D discrete nonpolynomial Schr\"odinger 
equation and 1D discrete Gross-Pitaevskii equation 
give the same results. 
On the other hand, if the on-site interaction strength is sufficiently 
large one finds deviations between 1D discrete nonpolynomial 
Schr\"odinger equation and 1D discrete Gross-Pitaevskii equation. 
By further increasing the attractive on-site interaction $U$ 
1D discrete Gross-Pitaevskii equation shows that eventually all the 
atoms accumulate into 
the same site. 1D discrete nonpolynomial Schr\"odinger equation 
shows instead something different: before 
all the atoms populate the same site there is the collapse 
of the condensate: 1D discrete nonpolynomial Schr\"odinger equation 
does not admit anymore a finite ground-state solution.

Numerically we find that the collapse occurs when $U<0$ and 
\beq 
{|U|N\over J} > 2.1 
\eeq
which is consistent with analytical result 
${|U|N/J} > 8/3$ of the continuum limit \cite{sala-npse}. 

\section{Dimensional reduction of a continuous quantum field theory}

A full quantum treatment of interacting bosons in a optical 
lattice is obtained by promoting the wavefunction $\psi({\bf r},t)$ 
of the 3D Gross-Pitaevskii equation (\ref{3dgpe}) to a field operator 
${\hat \psi}({\bf r},t)$ \cite{sala-book}, namely 
\beqa 
\psi(\mathbf{r},t) \to {\hat \psi}(\mathbf{r},t) \; , 
\\
\psi^*(\mathbf{r},t) \to {\hat \psi}^+(\mathbf{r},t) \; . 
\eeqa
The bosonic field operator 
${\hat \psi}({\bf r},t)$ and its adjunct ${\hat \psi}^+({\bf r},t)$
must satisfy the following equal-time commutation rules 
\beq
[ {\hat \psi}({\bf r},t) , {\hat \psi}^+({\bf r}',t) ] =
\delta({\bf r}-{\bf r}') \; , 
\eeq
\beq
[ {\hat \psi}({\bf r},t) , {\hat \psi}({\bf r}',t) ] = 
[ {\hat \psi}^+({\bf r},t) , {\hat \psi}^+({\bf r}',t) ] = 0 \; , 
\label{commuta-s-bose}
\eeq
By imposing these commutation rules one finds 
\beq 
{\hat \psi}^+({\bf r},t) |0\rangle = |{\bf r},t\rangle \; , 
\eeq
that is the operator ${\hat \psi}^+({\bf r},t)$ 
creates a particle in the state $|{\bf r},t\rangle$ from 
the vacuum state $|0\rangle$, and also 
\beq 
{\hat \psi}({\bf r},t) |{\bf r}'t\rangle = \delta({\bf r}-{\bf r}') \ 
|0\rangle \; , 
\eeq
that is the operator ${\hat \psi}({\bf r},t)$ 
annihilates a particle which is in the state $|{\bf r},t\rangle$. 

After promoting the wavefunction $\psi(\mathbf{r},t)$ to a field operator 
${\hat \psi}(\mathbf{r},t)$, Eq. (\ref{3dgpe}) becomes 
\beqa
i{\frac{\partial}{\partial t}}{\hat \psi}(\mathbf{r},t) &=&
\Big[ -{\frac{1}{2}}\nabla^{2}
+ {1\over 2} \left( x^2 + y^2 \right)
+ V(z) 
\nonumber 
\\
&+& 2\pi g {\hat \psi}^+(\mathbf{r},t)
{\hat \psi}(\mathbf{r},t) \Big] {\hat \psi}(\mathbf{r},t) \; , 
\label{quantum-3dgpe}
\eeqa
This equation is nothing else than the Heisenberg equation 
of motion 
\beq 
i {\partial \over \partial t} {\hat \psi} 
= [ {\hat \psi} , {\hat H} ] \;  
\eeq
of the field operator ${\hat \psi}(\mathbf{r},t)$, where 
\beq 
{\hat H} = \int d^3{\bf r} \ 
{\hat \psi}^+
\left[ -{\frac{1}{2}}\nabla^{2}
+ {1\over 2} \left( x^2 + y^2 \right)
+ V(z) +2\pi g {\hat \psi}^+
{\hat \psi} \right] {\hat \psi}
\label{q3dgpe}
\eeq 
is the many-body quantum Hamiltonian of the system, 
which is not necessarily a BEC \cite{sala-book}. Thus, 
the many-body Hamiltonian (\ref{q3dgpe}) describes 
a dilute gas of bosonic atoms 
confined in the plane $(x,y)$ by the transverse harmonic potential 
and by a generic potential $V(z)$ in the $z$ direction. 

\subsection{Dimensional reduction of the Hamiltonian}

To perform the dimensional reduction of the Hamiltonian (\ref{q3dgpe}) 
we suppose that 
\beq
{\hat \psi}({\bf r}) |G\rangle = 
{1\over \pi^{1/2} \sigma(z)}
\exp{\left[ - \left( {x^2+y^2\over 2\sigma(z)^2} 
\right) \right] }\, {\hat \phi}(z) |G\rangle \; , 
\label{qassume}
\eeq
where $|G\rangle$ is the many-body ground state, while 
$\sigma(z)$ and ${\hat \phi}(z)$ account respectively 
for the transverse width and for the axial bosonic 
field operator. We apply this ansatz to Eq. (\ref{q3dgpe}) and obtain 
\beq 
{\hat H}|G\rangle = {\hat H}_{e} |G\rangle 
\eeq
where, neglecting the space derivatives of $\sigma(z)$, the 
effective 1D Hamiltonian reads 
\beq
{\hat H}_{e} = \int dz \ {\hat \phi}^+ 
\Big[ -{\frac{1}{2}}\partial_z^{2} + V(z) + {1\over 2} 
\left( {1\over \sigma^2} + \sigma^2 \right) 
+ {g\over 2\sigma^2} \ {\hat \phi}^+  {\hat \phi} \Big] {\hat \phi} \; .  
\label{qe1}
\eeq
The transverse width $\sigma(z)$ can be determined 
by averaging the Hamiltonian (\ref{qe1}) over the 
ground state $|G\rangle$ and minimizing the resulting energy functional  
\beqa 
\langle G|{\hat H}_{e}|G\rangle 
&=& \int dz \Big\{ 
\langle G| {\hat \phi}^+ 
\Big[ -{\frac{1}{2}}\partial_z^{2} + V(z) \Big] {\hat \phi} |G\rangle
\nonumber
\\
&+& {1\over 2} 
\left( {1\over \sigma^2} + \sigma^2 \right) 
\langle G| {\hat \phi}^+ {\hat \phi} |G\rangle
\nonumber 
\\
&+& {g\over 2\sigma^2} 
\langle G| {\hat \phi}^+   {\hat \phi}^+ {\hat \phi} 
{\hat \phi} |G\rangle \Big\} 
\label{functional}
\eeqa
with respect to $\sigma(z)$. In this way one gets 
\beq 
\sigma(z)^4 =  
1 + g 
{
\langle G|{\hat \phi}^+(z){\hat \phi}^+(z)
{\hat \phi}(z){\hat \phi}(z)|G\rangle 
\over 
\langle G|{\hat \phi}^+(z){\hat \phi}(z)|G\rangle
}
\label{qsig1}
\; . 
\eeq
Thus, the ground state $|G\rangle$ is obtained self-consistently 
from Eqs. (\ref{qe1}) and (\ref{qsig1}). 
Notice that introducing the local axial-density operator 
${\hat \rho}(z) = {\hat \phi}^+(z) {\hat \phi}(z)$, such that 
$\rho(z) = \langle G| {\hat \phi}^+(z){\hat \phi}(z)|G \rangle= 
\langle G| {\hat \rho}(z)|G \rangle$ is the local axial density 
and $\rho_2(z)= 
\langle G | {\hat \phi}^+(z)  {\hat \phi}^+(z) {\hat \phi}(z) 
{\hat \phi}(z) = \langle 
G | {\hat \rho}(z){\hat \rho}(z)|G \rangle  - 
\delta(0)\, \rho(z)$ is the two-body axial correlation function, 
Eq. (\ref{qsig1}) can be rewritten as 
\beq 
\sigma(z)^4 = 1 + g{\rho_2(z)\over \rho(z)} \; . 
\label{sig2}
\eeq
Clearly, if $g\rho_2(z)\ll \rho(z)$ one has 
\beq 
\sigma=1 
\eeq
and the effective Hamiltonian (\ref{qe1}) reduces to 
\beq 
{\hat H}_e = {\hat H}_{1D} + 1 \; , 
\eeq
where ${\hat H}_{1D}$ is the strictly one-dimensional Hamiltonian 
\beq 
{\hat H}_{1D} = \int dz \ {\hat \phi}^+ 
\Big[ -{\frac{1}{2}}\partial_z^{2} + V(z) 
+ {g\over 2} \ {\hat \phi}^+  {\hat \phi} \Big] {\hat \phi} \;   
\eeq
while $1$ is the transverse energy (in units of $\hbar \omega_{\bot}$). 

Let us analyze the general case $\sigma(z) \neq 1$. 
In the superfluid regime, where $|G\rangle$ 
is the Glauber coherent state $|GCS\rangle$ of ${\hat \phi}(z)$ 
\cite{sala-book}, i.e. such that 
\beq 
{\hat \phi}(z) |GCS\rangle = \phi(z) |GCS\rangle \; , 
\eeq
from Eq. (\ref{qsig1}) one finds 
\beq 
\sigma(z)^4 = 1 + g |\phi(z)|^2 \;   
\eeq
and the energy functional (\ref{functional}) then becomes 
\beq 
\langle GCS|{\hat H}_{e}|GCS\rangle 
= \int dz \ 
\phi^* \Big[ -{1\over 2} \partial_z^{2} + V(z) 
+ \sqrt{1 + g |\phi|^2} \Big] \phi \; .  
\label{functional-npse} 
\eeq
This is the familiar energy functional of the 1D nonpolynomial Schr\"odinger 
equation \cite{sala-npse}.

\subsection{1D nonpolynomial Heisenberg equation}

From the effective 1D Hamiltonian (\ref{qe1}), the Heisenberg equation 
of motion 
\beq 
i {\partial \over \partial t} {\hat \phi} 
= [ {\hat \phi} , {\hat H}_e ] \;  
\eeq 
gives 
\beqa 
i {\partial \over \partial t} {\hat \phi}(z,t) &=& 
\Big[ -{\frac{1}{2}}\partial_z^{2} + V(z) + {1\over 2} 
\left( {1\over \sigma^2(z,t)} + \sigma^2(z,t) \right) 
\nonumber 
\\
&+& {g\over \sigma(z,t)^2} \ {\hat \phi}^+(z,t) {\hat \phi}(z,t) \Big] 
{\hat \phi}(z,t) \; , 
\eeqa
that is a 1D nonpolynomial Heisenberg equation because 
it must be solved self-consistently with the equation 
\beq 
\sigma(z,t)^4 =  
1 + g {
\langle S|{\hat \phi}^+(z,t){\hat \phi}^+(z,t)
{\hat \phi}(z,t){\hat \phi}(z,t)|S\rangle 
\over 
\langle S|{\hat \phi}^+(z,t){\hat \phi}(z,t)|S\rangle
}
\label{sig1t}
\; ,  
\eeq
where $|S\rangle$ is the many-body quantum state on the system. 
Only if the many-body state $|S\rangle$ coincides with the 
Glauber coherent state $|GCS\rangle$ \cite{sala-book}, such that 
${\hat \phi}(z,t)|GCS\rangle = \phi(z,t)|GCS\rangle$, the 
1D nonpolynomial Heisenberg equation reduces 
to the 1D nonpolynomial Schrodinger equation 
\cite{sala-npse}, given by 
\beqa 
i {\partial \over \partial t} \phi(z,t) &=& 
\Big[ -{\frac{1}{2}}\partial_z^{2} + V(z) + {1\over 2} 
\left( {1\over \sigma^2(z,t)} + \sigma^2(z,t) \right) 
\nonumber 
\\
&+& {g\over \sigma(z,t)^2} \ |\phi(z,t)|^2 \Big] \phi(z,t) \; , 
\eeqa
where $\phi(z,t)$ is a complex wavefunction and 
\beq 
\sigma(z,t) = \left( 1 + g |\phi(z,t)|^2 \right)^{1/4} \;  
\eeq
is the corresponding transverse width. 

\subsection{Generalized Lieb-Liniger theory}

In the time-independent and uniform case, where $V(z)=0$, the space-time  
dependence in Eq. (\ref{sig1t}) disappears, i.e.  
\beq 
\sigma^4 = 1 + g{\rho_2 \over \rho} \; , 
\label{sig3}
\eeq 
and the energy functional (\ref{functional}) reduces to a function 
of $\rho$, $\rho_2$ and $\sigma$, namely 
\beq
\langle G|{{\hat H}_{e}\over L}|G\rangle 
= \langle G| {\hat \phi}^+ 
\Big[ -{\frac{1}{2}}\partial_z^{2} \Big] {\hat \phi} |G\rangle 
+ {g \over 2\sigma^2} \rho_2 
+ {1\over 2} 
\left( {1\over \sigma^2} + \sigma^2 \right) \rho \; ,    
\label{function}
\eeq
where $L$ is the length of the uniform system. Due to the Lieb-Liniger 
theorem \cite{lieb}, for $g\geq 0$ the energy function (\ref{function}) 
can be rewritten as 
\beq
\langle G|{{\hat H}_{e}\over L}|G\rangle 
= {1\over 2} \rho^3 f({g\over \rho \sigma^2}) + {1\over 2} 
\left( {1\over \sigma^2} + \sigma^2 \right) \rho \; ,   
\label{lieb}
\eeq
where $f(x)$ is the Lieb-Liniger function, which is defined 
as the solution of a Fredholm equation and it is such that 
$f(x)=x-4x^{3/2}/(3\pi)$ for $x\ll 1$ and $f(x)=(\pi^2/3)(x/(x+2))^2$ 
for $x\gg 1$. The minimization of (\ref{lieb}) with respect 
to $\sigma$ gives  
\beq 
\sigma^4 = 1 + g \rho \ f'({g\over \rho \sigma^2}) \; , 
\label{gallinavecchia}
\eeq
and consequently, comparing with Eq. (\ref{sig3}), the two-body 
axial correlation function $\rho_2$ must satisfy the equation 
\beq 
\rho_2 = \rho^2 \ f'({g\over \rho \sigma^2}) \; . 
\eeq
Notice that Eqs. (\ref{lieb}) and (\ref{gallinavecchia}), which are 
a reliable generalization of the Lieb-Lineger theory, have been 
obtained for the first time by Salasnich, Parola and 
Reatto \cite{sala-lieb1} using a many-orbitals variational approach. 
As discussed in the introduction, some years ago we used this generalized 
Lieb-Liniger theory to analyze the transition 
from a 3D Bose-Einstein condensate to the 1D Tonks-Girardeau gas 
\cite{sala-lieb1}, showing that 
the experimental data on a Tonks-Girardeau gas of $^{87}$Rb atoms 
of Kinoshita, Wenger, and Weiss \cite{kinoshita} are very well 
described by our theory that takes into account 
variations in the transverse width of the 
atomic cloud \cite{sala-lieb2}. 

\section{Dimensional reduction for bosons in a quasi-1D lattice}

To conclude this chapter, we perform a discretization 
of the 3D many-body Hamiltonian (\ref{q3dgpe}) along the $z$ 
axis due to the presence of the periodic potential, given by 
Eq. (\ref{periodic}). We use the decomposition \cite{book-lattice}
\beq
{\hat \psi}(\mathbf{r}) = \sum_n {\hat \phi}_n(x,y) \ W_n(z) \; , 
\eeq
that is the quantum-field-theory analog of Eq. (\ref{zanno}) and 
we set up the quantum-field-theory 
extension of the mean-field approach developed in the first 
part of this contribution. In particular we write 
\beq
{\hat \phi}_n(x,y) |G\rangle = 
{1\over \pi^{1/2} \sigma_n}
\exp{\left[ - \left( {x^2+y^2\over 2\sigma_n^2} 
\right) \right] }\, {\hat b}_n |G\rangle \; , 
\eeq
where $|G\rangle$ is the many-body ground state,  while 
$\sigma_n$ and ${\hat b}_n$ account respectively 
for the on-site transverse width and for the bosonic 
field operator. We insert these ansatz into Eq. (\ref{q3dgpe}) and 
we easily obtain the effective 1D Bose-Hubbard Hamiltonian \cite{sala-barba}
\beqa 
{\hat H}_e &=& \sum_{n} \Big\{ 
\big[ {1\over 2} ( {1\over \sigma_n^2} +  
\sigma_n^2) + \epsilon_n \big] {\hat n}_n 
- J \, {\hat b}_n^+ \left( {\hat b}_{n+1}+{\hat b}_{n-1} \right) 
\nonumber 
\\
&+& {1\over 2} {U \over \sigma_n^2} {\hat n}_n ({\hat n}_n-1) \Big\} \; . 
\label{ee1}
\eeqa
where ${\hat n}_n={\hat b}_n^+{\hat b}_n$ is the on-site number operator, 
$\epsilon_n$ is the on-site axial energy, while $J$ and $U$ 
are the familiar hopping (tunneling) energy and on-site energy, 
given by Eqs. (\ref{maroni1}), (\ref{maroni2}) and (\ref{maroni3}). 

Our Eq. (\ref{ee1}) takes into account deviations with respect 
to the strictly 1D case 
due to the transverse width $\sigma_n$ of the bosonic field. 
This on-site transverse width $\sigma_n$ can be determined 
by averaging the Hamiltonian (\ref{ee1}) over a many-body 
quantum state $|G\rangle$ and minimizing the resulting energy function 
with respect to $\sigma_n$. In this way one gets \cite{sala-barba}
\beq
\sigma_n^4 = 1 + U {\langle G|{\hat n}_n^2|G\rangle - 
\langle G|{\hat n}_n |G \rangle 
\over \langle G | {\hat n}_n |G \rangle } \; . 
\label{ssig1} 
\eeq
Note that Eqs. (\ref{ee1}) and (\ref{ssig1}) must be solved 
self-consistently to obtain the ground-state of the system. 
Clearly, if $U<0$ the transverse width $\sigma_n$ is smaller than 
one (i.e. $\sigma_n <a_{\bot}$ in dimensional units) and 
the collapse happens when $\sigma_n$ goes to zero. 
At the critical strength $U_c$ of the collapse 
all particles are accumulated in few sites 
and consequently $U_c \simeq -1/N$. 

We stress that, from Eq. (\ref{ssig1}), the system is strictly 
1D only if the following strong inequality 
\beq 
U  {\langle {\hat n}_n^2\rangle - \langle {\hat n}_n\rangle 
\over \langle {\hat n}_n \rangle }  \ll 1  
\label{condition}
\eeq 
is satisfied for any $n$, such that $\sigma_n=1$ 
(i.e. $\sigma_n=a_{\bot}$ in dimensional units). Under 
the condition (\ref{condition}) the problem of collapse 
is fully avoided. In this strictly 1D regime where 
the effective Hamiltonian of Eq. (\ref{ee1}) becomes (neglecting 
the irrelevant constant transverse energy) 
\beq 
{\hat H}_{1D} = \sum_n \epsilon_n {\hat n}_n 
-J \sum_n  {\hat b}_n^+ \left( {\hat b}_{n+1}+{\hat b}_{n-1} \right) 
+\frac{U}{2} \sum_n {\hat n}_n ({\hat n}_n-1) 
\eeq
which is the familiar 1D Bose-Hubbard model \cite{book-lattice}. 

Given the generalized Bose-Hubbard Hamiltonian (\ref{ee1}), 
the discrete Heisenberg equation of motion of the bosonic operator 
${\hat b}_n$ reads 
\beq 
i {\partial \over \partial t}{\hat b}_n = [{\hat b}_n , {\hat H}_e] \; ,  
\eeq
that is 
\beq 
i {\partial \over \partial t}{\hat b}_n = 
\big[ {1\over 2} ( {1\over \sigma_n^2} +  
\sigma_n^2) + \epsilon_n \big] {\hat b}_n 
- J \, \left( {\hat b}_{n+1}+{\hat b}_{n-1} \right) 
+ {U \over \sigma_n^2} {\hat n}_n \ {\hat b}_n \; . 
\eeq
This is a 1D discrete nonpolynomial Heisenberg equation because 
it must be solved self-consistently with the equation 
\beq
\sigma_n^4 = 1 + U {\langle S|{\hat n}_n^2|S\rangle - 
\langle S| {\hat n}_i |S\rangle 
\over \langle S|{\hat n}_n |S\rangle }\; . 
\eeq
where $|S\rangle$ is the many-body quantum state on the system. 
Also in this discrete case, only if the many-body state 
$|S\rangle$ coincides with the 
Glauber coherent state $|GCS\rangle$, such that 
\beq 
{\hat b}_n \ |GCS\rangle = f_n \ |GCS\rangle \; , 
\eeq
the 1D discrete nonpolynamial Heisenberg equation reduces 
to the 1D discrete nonpolynomial Schr\"odinger 
equation, given by Eqs. (\ref{e1}) and (\ref{sig1}). 

\section{Conclusions}

We have investigated the discrete bright solitons of a quasi-one-dimensional 
Bose-Einstein condensate with axial periodic potential 
by using an effective one-dimensional 
discrete nonpolynomial Schr\"odinger equation \cite{sala-dnpse1,sala-dnpse2}. 
We have shown that, contrary to the familiar  one-dimensional 
discrete nonlinear Schr\"odinger equation, our 
gives rise to the collapse of the condensate above a 
critical (attractive) strength, in agreement with experimental data. 
We have also analyzed the dimensional reduction 
of a bosonic quantum field theory finding an effective 1D quantum 
Hamiltionian (and a corresponding effective 1D nonpolynomial Heisenberg 
equation) which gives a generalized Lieb-Liniger theory 
in the absence of axial periodic potential 
\cite{sala-lieb1,sala-lieb2} and gives instead a generalized 
Bose-Hubbard model \cite{sala-barba} in the presence 
of axial periodic potential. In Ref. \cite{sala-barba} we have used 
the Density-Matrix-Renormalization-Group 
(DMRG) technique to study the bright solitons of the 1D Bose-Hubbard 
Hamiltonian finding that beyond-mean-field 
effects become relevant by increasing the attraction between 
bosons. In particular we have discover that, contrary to the MF predictions 
based on the discrete nonlinear Schr\"odinger equation, 
quantum bright solitons are not self-trapped \cite{sala-barba}.  
In other words, we have found that with a 
small number $N$ of bosons the average of the quantum
density profile, that is experimentally obtained with
repeated measures of the atomic cloud, is not shape
invariant. This remarkable effect can be explained by 
considering a quantum bright soliton as a MF bright 
soliton with a center of mass that is randomly 
distributed due to quantum fluctuations, which are
suppressed only for large values of $N$ \cite{sala-barba}.

\section*{Acknowledgments}

The author acknowledges for partial support 
Universit\`a di Padova (Progetto di Ateneo), 
Cariparo Foundation (Progetto di Eccellenza), 
and MIUR (Progetto PRIN). 
The author thanks L. Barbiero, B. Malomed, A. Parola, 
V. Penna, and F. Toigo for fruitful discussions. 

\newcommand{\noopsort}[1]{} \newcommand{\printfirst}[2]{#1}
\newcommand{\singleletter}[1]{#1} \newcommand{\switchargs}[2]{#2#1}

\bibliographystyle{spmpsci}

\begin{thebibliography}{2}

\bibitem{sala-barba} Barbiero, L., Salasnich, L.: 
Quantum bright solitons in a quasi-one-dimensional optical lattice. 
Phys. Rev. A {\bf 89}, 063605 (2014)

\bibitem{bloch_review} Bloch, I., Dalibard, J., Zwerger, W.: 
Many-body physics with ultracold gases. Rev. Mod. Phys. {\bf 80}, 885 (2008)

\bibitem{cazalilla2011} Cazalilla, M.A., Citro, R., 
Giamarchi, T., Orignac, E., Rigol, M.: 
One dimensional Bosons: From Condensed Matter Systems to Ultracold Gases. 
Rev. Mod Phys. {\bf 83} 1405 (2011) 

\bibitem{sala-numerics} Cerboneschi, E., Mannella, R., Arimondo, E., 
Salasnich, L.: Oscillation Frequencies for a Bose Condensate 
in a Triaxial Magnetic Trap. Phys. Lett. A {\bf 249}, 495 (1998); 
Mazzarella, G., Salasnich, L.: Collapse of triaxial bright solitons 
in atomic Bose-Einstein condensates. Phys. Lett. A {\bf 373}, 4434 (2009)

\bibitem{exp-solo3} Cornish, S.L., Thompson, S.T., Wieman, C.E., 
Formation of bright matter-wave solitons during the collapse 
of attractive Bose-Einstein condensates. 
Phys. Rev. Lett. {\bf 96}, 170401 (2006)

\bibitem{exp-gap} Eiermann, B., Anker, Th., Albiez, M., 
Taglieber, M., Treutlein, P., Marzlin, K.P., Oberthaler, M.K.:  
Bright Bose-Einstein Gap Solitons of Atoms with Repulsive Interaction.  
Phys. Rev. Lett. {\bf 92}, 230401 (2004) 

\bibitem{giamarchi} Giamarchi, T.: Quantum Physics in One Dimension. 
Oxford Univ. Press, Oxford (2004)

\bibitem{bloch} M. Greiner, O. Mandel, T. Esslinger, T H\"ansch, 
and I. Bloch, Nature {\bf 415}, 39 (2002).

\bibitem{book-bose} Leggett, A.J.: Quantum Liquids. 
Bose condensation and Cooper Pairing in Condensed-Matter Systems.  
Oxford Univ. Press, Oxford (2006) 

\bibitem{book-lattice}  Lewenstein, M., Sanpera, A., Ahufinger, V.: 
Ultracold Atoms in Optical Lattices: 
Simulating Quantum Many-Body Systems. Oxford Univ. Press, 
Oxford (2012) 

\bibitem{lieb} E.H. Lieb and W. Liniger, 
Exact Analysis of an Interacting Bose Gas. I. 
The General Solution and the Ground State. Phys. Rev. {\bf 130}, 1605 (1963)

\bibitem{sala-dnpse1} A. Maluckov, L. Hadzievski, B.A. Malomed, 
and L. Salasnich, Solitons in the discrete nonpolynomial 
Schr\"odinger equation. Phys. Rev. A {\bf 78}, 013616 (2008)

\bibitem{sala-dnpse2} G. Gligoric, A. Maluckov, L. Salasnich, B.A. Malomed, 
and L. Hadzievski, Two routes to the one-dimensional discrete 
nonpolynomial Schr\"dinger equation. Chaos {\bf 19}, 043105 (2009) 

\bibitem{exp-solo4} Marchant, A.L., Billam, T.P., 
Wiles, T.P., Yu, M.M.H., Gardiner S.A., Cornish, S.L.: 
Controlled formation and reflection of a bright solitary matter-wave. 
Nat. Commun. {\bf 4}, 1865 (2013)

\bibitem{morsch} Morsch, O., Oberthaler, M.:
Dynamics of Bose-Einstein condensates in optical lattices. 
Rev. Mod. Phys. {\bf 78}, 179 (2006)

\bibitem{kevre} Kevrekidis, P.G.: 
The Discrete Nonlinear Schrodinger Equation: Mathematical Analysis, 
Numerical Computations and Physical Perspectives
(Springer, New York, 2009). 

\bibitem{exp-solo1} Khaykovich, L., Schreck, F., Ferrari, F.G., 
Bourdel, T., Cubizolles, J., Carr, L.D., Castin, Y., Salomon, C.:  
Formation of a Matter-Wave Bright Soliton. Science {\bf 296}, 1290 (2002)

\bibitem{kinoshita} Kinoshita, T., Wenger, T., Weiss, D.S.: 
Observation of a one-dimensional Tonks-Girardeau gas. 
Science {\bf 305}, 1125 (2004)

\bibitem{sala-npse} Salasnich, L.: Pulsed Quantum Tunneling with Matter Waves. 
Laser Physics {\bf 12}, 198(2002); Salasnich, L. Parola, A., Reatto, L., 
Effective wave-equations for the dynamics of cigar-shaped and 
disc-shaped Bose condensates. Phys. Rev. A {\bf 65}, 043614 (2002)

\bibitem{sala-book} Salasnich, L.: Quantum Physics of Light and Matter.  
A Modern Introduction to Photons, Atoms and Many-Body Systems.  
Springer, Cham (2014) 

\bibitem{sala-gaussian} Salasnich, L.: 
Time-dependent variational approach to Bose-Einstein condensation. 
Int. J. Mod. Phys. B {\bf 14}, 1 (2000)

\bibitem{sala-lieb1} Salasnich, L., Parola, A., Reatto, L.: 
Transition from 3D to 1D in Bose Gases at Zero Temperature. 
Phys. Rev. A {\bf 70}, 013606 (2004) 

\bibitem{sala-lieb2} Salasnich, L., Parola, A., Reatto, L.: 
Quasi One-Dimensional Bosons in Three-dimensional Traps: From Strong 
Coupling to Weak Coupling Regime. Phys. Rev. A {\bf 72}, 025602 (2005)

\bibitem{exp-solo2} Strecker, K.E., Partridge, G.B., 
Truscott, A.G. and Hulet, R.G.: 
Formation and propagation of matter-wave soliton trains.
Nature (London) {\bf 417}, 150 (2002)

\end{thebibliography}

\end{document}